\documentclass[preprint2]{aastex}

\usepackage{epsfig}
\usepackage{amsmath}

\slugcomment{Submitted to ApJ}

\shorttitle{Salmonson and Galama}
\shortauthors{Kinematics of GRBs and Afterglows}

\begin{document}


\title{Discovery of a tight correlation between pulse lag/luminosity
and jet-break times: a connection between gamma-ray burst and
afterglow properties. }


\author{Jay D. Salmonson} 
\affil{Lawrence Livermore National Laboratory, L-095, P.O. Box 808, Livermore, CA 94551}

\author{Titus J. Galama}
\affil{Division of Physics, Mathematics and Astronomy, 105-24, 
California Institute of Technology, Pasadena, CA 91125}



\begin{abstract}

A correlation is presented between the pulse lag and the jet-break
time for seven BATSE gamma-ray bursts with known redshifts.  This is,
to our best knowledge, the first known direct tight correlation
between a property of the gamma-ray burst phase (the pulse lag) and
the afterglow phase (the jet-break time). As pulse lag and luminosity
have been found to be correlated this also represents a correlation
between peak luminosity and jet-break time.  Observed timescales
(variability or spectral lags) as well as peak luminosity naturally
have a strong dependence on the Lorentz factor of the outflow and so
we propose that much of the variety among GRBs has a purely kinematic
origin (the speed or direction of the outflow).

We explore a model in which the variation among GRBs is due to a
variation in jet-opening angles, and find that the narrowest jets have
the fastest outflows. We also explore models in which the jets have
similar morphology and size, and the variation among bursts is caused
by variation in viewing angle and/or due to a velocity profile. The
relations between luminosity, variability, spectral lag and jet-break
time can be qualitatively understood from models in which the Lorentz
factor decreases as a function of angle from the jet axis. One expects
to see high luminosities, short pulse lags and high variability as
well as an early jet-break time for bursts viewed on axis, while
higher viewing inclinations will yield lower luminosities, longer
pulse lags, smoother bursts and later jet-break times.

\end{abstract}




\section{Introduction }

Pulse peaks in Gamma-Ray Burst (GRB) lightcurves evolve in time from
higher to lower energies and become wider. \citet{nmb00} provide
measurements of such lags as observed in GRBs between different energy
channels of the Burst And Transient Source Experiment (BATSE). The
lags are measured by calculating the cross-correlation function
(CCF). In a sample of GRBs for which the redshift is known it is found
that the spectral lag is related to the isotropic gamma-ray peak
luminosity \citep{nmb00}. Similarly \citet{fr00} have shown that
measures of variability in GRB lightcurves (see also Reichart et
al. [2001]\nocite{rlfr+01}) are related to the isotropic gamma-ray peak
luminosity. Vice versa, measurements of variability or of spectral
lags can then be used as a crude distance indicator for the GRB.

Observed timescales (variability or spectral lags) as well as peak
luminosity naturally have a strong dependence on the Doppler factor of
the outflow (a function of the Lorentz factor and the direction of 
motion with respect to the observer). If indeed the Doppler factor is
the dominant parameter among GRBs, a relation between spectral
lags/variability and luminosity is expected.

Crucial to our understanding of what causes the GRB is the question of
whether GRB engines are in some sense ``standard
candles''. Observationally it is found that the isotropic equivalent
energies of GRBs range from about 5 $\times$ 10$^{51}$ to 1.4 $\times$
10$^{54}$ ergs \citep{bfs01}. However, transitions have been observed
at optical and radio wavelengths which can be interpreted as being due
to collimated (jetted) outflow \citep{hbf+99}.  When correcting the
observed $\gamma$-ray energies for the geometry of the outflow, GRB
energies appear narrowly clustered around $5 \times 10^{50}$ ergs
(Frail et al. 2001; see also Panaitescu and Kumar
2001\nocite{fksd01,pk01}. \citet{fksd01} infer the jet opening angle
from the observed jet-break time and find that there is a wide range
in opening angles. The reason for why this range in angles exists is
currently not understood.

The idea that GRB energies may be narrowly clustered was also put
forward by \citet{jay00,jay01}. However, \citet{jay01} proposed that there
exists not a range in opening angles, but that all bursts derive from
a single-burst-jet morphology and that the variation in viewing angle
of the jet yields the observed variation among GRBs.

In this paper is presented a tight correlation between jet-break
times, $\tau_j$, and pulse lags, $\Delta t$. Since luminosity and
pulse lags have been shown to be correlated, this also represents a
relation between luminosity and jet-break times. As time scales and
peak luminosity are strong functions of the Doppler factor of the
outflow, this relation suggests the jet-break time to also be a strong
function of the Doppler factor. We discuss the implications of this
result on our understanding of the morphology of the explosion.

\section{Pulse Lag and Luminosity versus Jet-Break Times  } \label{datasection}

The relativistic blast-wave model has become the `standard' model for
the interpretation of GRBs and their afterglows (see Piran
[2000]\nocite{pira00} for a review).  It invokes the release of a
large amount of energy, resulting in an ultra-relativistic outflow. In
this model the GRB is produced in internal shocks. During or after the
GRB emission phase a strong shock is formed when these (merged) shells
run into the surrounding medium. As the (forward) shock is weighed
down by increasing amounts of swept-up material it produces a slowly
fading `afterglow' at radio to X-ray wavelengths. Observations of the
GRB and the afterglow at gamma- and X-ray wavelengths \citep{gvk+99}
suggest that indeed in some cases the afterglow commences during or
shortly after the GRB.

If the GRB is due to internal shocks and the afterglow is due to the
external forward shock then little connection between the two
phenomena is to be expected. Nevertheless the afterglow is the result
of the interaction of colliding shells (that produced the GRB) with
the ambient medium and some level of imprint from the GRB phase may be
expected to be present in the afterglow. 

The clustering of GRB energies (a property of the GRB phase) inferred
from measurements of the jet-break times (a property of the afterglow)
suggests a possible relation between GRB and afterglow properties.  We
investigate a complete sample of seven GRBs for which measurements of
the redshift, the spectral lag, the luminosity, and of the jet-break
time are available. We collect redshifts and jet-break times from
\citet{fksd01}, spectral data from \citet{jbp01} and BATSE pulse lags
for six bursts from \citet{nmb00}. To this set we have added the data
of GRB\,991216 (observed lag $\Delta t_{\text{obs}} =
0.018^{+0.002}_{-0.001}$ seconds; J.\ P.\ Norris 2001, private
communication).

The correlation between rest-frame BATSE pluse lags, $\Delta t$, and
jet-break times, $\tau_j \equiv t_j/(1 + z)$, is shown in Figure
\ref{tjvtlag}; bursts with longer lags have longer jet-break
times. The data is fit with a least-squares fit using errors in
two-dimensions where errors on spectral lags are given in
\citet{nmb00} and errors in jet-break time are considered presently.

Jet-break times from the sample of \citet{fksd01} have not been
determined uniformly: some of the jet-break times are derived from
multi-band modeling (covering the radio to X-ray passbands), while
other events have primarily been observed in a single passband, and
are less well constrained. \citet{fksd01} estimate that uncertainties
in jet-break times range between 10-30\%. We therefore assume a 10\%
error for break times determined from broad-band modeling
(GRB\,990510), and 30\% for break times determined from mainly fitting
to a single passband (GRB\,970828, GRB\,990123, and GRB\,991216). We
make exceptions for GRB\,970508 and GRB\,980703 as follows. Frail,
Waxman \& Kulkarni (2000\nocite{fwk00}) have modeled GRB 970508 with a
wide-angled jet with $t_j$ = 25 days. The radio observations require a
jet-break but optical observations do not require such a transition,
though they are consistent with a late breaking at $t_j \sim 25$ days
\citep{rhoads99}. We therefore assume a large 50\% error in the
jet-break time of GRB\,970508 and caution the reader to consider this
number uncertain. GRB\,980703 has one of the richest multi-band data
sets. However, due to the brightness of its host galaxy the jet-break
time is not that well constrained, and multi-frequency modeling
requires a jet-break between $2-8$ days (E.Berger, private
communication; Berger et al. 2002, in preparation); we use $t_j = 5
\pm 3$ days.  For GRB\,971214 the jet-break time has not been measured
and \citet{fksd01} provided a lower limit. However, we are of the
opinion that the jet-break time of GRB\,971214 is essentially
unconstrained in view of the recent evidence for hard electron energy
distributions in GRB afterglows \citep{pan01,bha01}. Such
distributions result in slowly decaying afterglows. In addition, the
optical light curve of GRB\,971214 is not very well sampled
\citep{ddc+98,kdr+98,hth+98,wg99}. These facts allow for a much
earlier jet-break transition than \citet{fksd01} have provided: it
could have occured between a few hours after the event and more than a
month after. We do not use GRB\,971214 in any of the following fits,
although for reference we show the jet-break time of \citet{fksd01} in
the figures. A small break time would be consistent with its close
proximity to the relations shown in the figures of this paper.

Using the CCF31 0.1 lags, determined by cross-correlating pulses in
BATSE channels 1 \& 3 down to 0.1 of the peak luminosity
\citep{nmb00}, a good fit results:
\begin{equation}
\tau_j = \frac{t_j}{1+z} = 28^{+ 18}_{- 11} 
{ \biggl(\frac{\Delta t_{\text{(CCF31 0.1)}} }{1~\text{sec}}\biggr)^{0.89 \pm 0.12} \text{days}}
\label{t_jeqn}
\end{equation} 
(shown in Fig. \ref{tjvtlag}) with a reduced chi-squared $\chi^2_r =
4.7/4$ and a respectable goodness-of-fit $Q = 0.31$ \citep{pftv88}.
With only seven GRBs (six used in the fit), the data is sparse, but
the correlation is clear and surprisingly tight.  \citet{nmb00} also
calculated lags fitting down to 0.5 of the peak luminosity which gives
a looser fit $\tau_j = 38^{+ 28}_{- 16} \Delta t_{\text{(CCF31
0.5)}}^{0.95 \pm 0.13}$ with $\chi^2_r = 6.2/4$ and a goodness-of-fit
$Q = 0.19$.  Note that Eqn.~(\ref{t_jeqn}) simply relates two
timescales and as such does not have a direct dependance on spectral
ambiguities involved in calculating k-corrections for the peak
luminosity.

Since pulse lag and luminosity are correlated \citep{nmb00},
transitivity suggests a third relation between luminosity and
jet-break times.  We calculate peak luminosites, $L_{pk}$, for the
seven GRBs, from BATSE peak fluxes and spectral parameters in
\citet{jbp01} and implement a k-correction as in \citet{bfs01}.  The
k-correction used here transforms the [50,300] keV luminosity band
observed at Earth to a [20,2000] keV band local to the burst.  Errors
in peak luminosity were calculated using the default systematic errors
used by \citet{bfs01}; in particular we take 10 \% errors in peak flux
and spectral break energy and 20 \% errors in the specral indices.  A
fit to the data is
\begin{equation}
L_{pk} = 28^{+ 6}_{- 5} \times 10^{51} \Biggl(\frac{\tau_j}{1~ \text{days}}\Biggr)^{-1.58 \pm 0.23} \text{ergs s}^{-1} \label{Leqn}
\end{equation}  
(shown in Fig.~\ref{Lvtlag}) with $\chi^2_r = 29/4$ and $Q \sim
10^{-6}$.  With the large $\chi^2_r$ this relation isn't nearly as
tight as that of Eqn.~(\ref{t_jeqn}), but it does highlight the trend
discovered by \citet{nmb00} that brighter bursts have shorter
timescales. We find that this trend applies for jet-break times,
whereas \citet{nmb00} found a similar trend for spectral lag. Note
also that the errors in the fitting parameters derived from the
chi-squared fit are not reliable due to the high $\chi^2_r$.

In \citet{fksd01} the isotropic energy, $E_{iso}$, is compared with
the jet-break time $\tau_j$.  As above, we fit these two quantities,
taking k-corrected values for $E_{iso}$ from \citet{bfs01}.  The fit
gives 
\begin{equation}
E_{iso} = 96^{+29}_{-22} \times 10^{51}\ \tau_j^{-1.97 \pm
0.36} \text{ergs} \label{Eeqn}
\end{equation}
(Fig.~\ref{Eisovtj}) with $\chi^2_r = 51/4$ and a goodness-of-fit $Q
\sim 10^{-10}$.  As is also the case for the relation shown in
Eqn.~(\ref{Leqn}), the quality of the fit is poor, but the trend that
bursts of higher energy have shorter jet-break timescales is plainly
evident.  This relation allows one to test the suggestion by
\citet{fksd01} that there is a reservoir for gamma-ray energy,
$E_\gamma$, that is constant for all GRBs.  For bursts with a large
jet-opening angle, $\theta_j$, this energy would be spread out,
resulting in lower fluxes and lower inferred isotropic energies,
$E_{iso}$. This suggests $E_\gamma \approx$ constant = $E_{iso}
\theta_j^2 \propto (E_{iso} \tau_j)^{3/4}$.  The assumption that energy
in gamma-rays is constant predicts $E_{iso} \propto \tau_j^{-1}$,
which is not consistent with the data.  Thus, while the range of
observed $E_\gamma$ is significantly smaller than that of $E_{iso}$,
it is not exactly constant.

The existence of these correlations, in particular the relation
between jet-break times and spectral lags, point to a surprisingly
direct relationship between the GRB phase (represented by the pulse
lag and the luminosity) and the afterglow phase (represented by the
jet break time).  In the next section we will argue that these
relationships are kinematic in origin.

\section{A Kinematic Connection } \label{modelsection}

The effectively linear relationship between the two timescales; pulse
lags $\Delta t$ and jet-break times $\tau_j$ (Eqn.~\ref{t_jeqn})
suggests a surprisingly direct connection between the GRB
phase, from which the pulse lags are derived, and the afterglow phase,
from which jet-break times derive.  An initial conclusion that can be
drawn is that, since the GRB phase is generally thought not to depend
on the density of the interstellar medium (ISM), this relation
disfavors explanations of the jet break deriving from external causes
such as sudden drops in ISM density \citep{kp00b}.  But what could be
the meaning of such a connection between the GRB and the
afterglow phase, expressed in the relations presented in
Figs.~\ref{tjvtlag}, \ref{Lvtlag} and \ref{Eisovtj}?

It was argued by \citet{jay00} that the spectral lag vs.\ isotropic
gamma-ray peak luminosity relationship \citep{nmb00,jay00} may have a
simple kinematic explanation.  Defining a Doppler factor
\begin{equation}
\begin{split}
{\cal D} &\equiv \frac{1}{\gamma (1 - \beta \cos \theta)(1+z)} \\
&\approx \frac{2 \gamma}{(1 + [\theta \gamma]^2) (1+z)} 
~ (\text{for}~ \theta \ll 1, \gamma \gg 1)
\label{deltaeqn}
\end{split}
\end{equation}
of the ejecta moving with angle $\theta$ from the line of sight, with
a velocity $\beta \equiv v/c$ at a redshift $z$, a proper GRB
timescale $\tau$ in the frame of the emitter will be observed as
\begin{equation}
 t =  \tau/{\cal D}  ~. \label{tdeltaeqn}
\end{equation}
To calculate the lab-frame luminosity we assume that the jet opening
angle is greater than the relativistic beaming angle $(\theta_j >
1/\gamma)$.  A given photon spectrum $\phi(E)$ (photons keV$^{-1}$
s$^{-1}$), which is relativistically invariant, will yield a peak
luminosity $L_{pk} \propto \int^{e_2}_{e_1} E \phi(E/{\cal D}) dE$ (ergs
s$^{-1}$) or peak number luminosity $N_{pk} \propto
\int^{e_2}_{e_1}\phi(E/{\cal D}) dE$ (photons s$^{-1}$) when viewed
through a lab-frame bandpass [$e_1,e_2$] \citep[e.g.,][]{fehk92}.  An
acceptable analytical fit to $\phi(E)$ is given as a broken power-law
\citep{bmf+93}. However, since most of the photon flux is from low
energies, we use, for the sake of argument, only the low end of the
power-law spectrum $\phi(E) \propto E^{-\alpha}$, keeping in mind that
there will be a correction of order $\log (E_0)$, which is small for
our purposes; here $E_0$ is the break energy in the energy spectrum
\citep{bmf+93}.  As such we find that the luminosities vary as:
\begin{equation}
L_{pk} \propto  {\cal D}^\alpha; ~N_{pk} \propto {\cal D}^\alpha ~.
\label{lumvsdeltaeqn}
\end{equation}
If spectral pulse-lag is due to some proper decay timescale $\Delta
\tau$, this time scale will be $\Delta t = \Delta \tau/{\cal D}$ in the
lab-frame. Observations find $\alpha \approx 1$ \citep{ppbm+98}. Thus,
combining Eqns.~(\ref{tdeltaeqn} \& \ref{lumvsdeltaeqn}), we argue
that the inverse relationship of the pulse lag - luminosity
correlation discovered by \citet{nmb00}
\begin{equation}
L_{pk} \propto \Delta t^{-1.14}
\label{jpnlageqn}
\end{equation}
and that discussed by \citet{jay00}
\begin{equation}
N_{pk} \propto \Delta t^{-0.98} ~,
\label{jdslageqn}
\end{equation}
result from a purely kinematic effect.  That is, faster expanding
bursts will appear to be brighter and to evolve more quickly than
slower bursts.

Assuming that spectral pulse lag is due to some proper timescale,
i.e. $\Delta t \propto 1/{\cal D}$, and using the relation of
Eq. \ref{t_jeqn}, we find that also the jet-break time $\tau_j \propto
1/{\cal D}$. Somehow the jet-break time depends on the initial
conditions (as reflected in the Doppler factor during the GRB phase),
and those conditions are maintained in the afterglow.

This suggests that the relations $L_{pk}(\tau_j)$ and
$E_{iso}(\tau_j)$ discussed in Section \ref{datasection} could be
manifestations of the same effect.  For sake of argument, if we assume
that $L_{pk}$ is constant over a burst duration, then we have $E_{iso}
= L_{pk} t_{tot}$, where $t_{tot}$ is the burst duration.  In our
sample, burst duration, which is typically $t_{tot} \sim 10$ seconds,
varies within a factor of $~2$ while $L_{pk}$ and $E_{tot}$ vary
over a factor $\sim 100$.  This makes sense because $t_{tot}$ is equal
to the lifetime of the central engine and thus is not affected by
kinematics, unlike $L_{pk}$ and $E_{tot}$.  Taking $t_{tot}$ to be
essentially constant compared to $L_{pk}$ and $E_{iso}$, then, to the
extent that $L_{pk}$ is constant over the burst duration, we expect
$E_{iso} \sim L_{pk}$.  The variability-luminosity relationship
\citep{rlfr+01} would imply that dimmer burts are less variable in
luminosity and thus we expect they will have the strongest correlation
$E_{iso} \sim L_{pk}$ (they come closest to the assumption of constant
luminosity over the burst duration), while brighter, more variable
bursts will appear more chaotic; a trend that appears to be reflected
in the data (see Figures \ref{Lvtlag} and \ref{Eisovtj}).  Thus,
overall, it is not surprising that both $L_{pk}$ and $E_{iso}$ have
similar slopes in $\tau_j$.

Given that observed timescales in general vary as $t \propto
\tau/{\cal D}$, and, as discussed above, luminosities also appear to
be functions of the Doppler boost, we are led to the intriguing
possibility that all of the relationships described here
(Eqns. \ref{t_jeqn}, \ref{Leqn}, \ref{jpnlageqn} \& \ref{jdslageqn})
depend on one kinematic variable: ${\cal D}$. Thus the existence of
the relationships between spectral lag, $\Delta t$, jet-break times,
$\tau_j$ and peak luminosity, $L_{pk}$ (or $N_{pk}$), would be
evidence for variation in motion among GRBs. We here propose that, in
fact, kinematic variation is the dominant source of variety observed
among GRBs.  The specific nature of this kinematic variation is still
unclear.  Specifically, is it the opening angle $\theta$ of the jet, or
is the Lorentz factor $\gamma$ the dominant source of variation in
${\cal D}$ (Eqn.~\ref{deltaeqn})? This question will be explored with
three possible models.

\subsection{Variation in the burst population: Many Morphologies \label{burstpop}}

This model was put forward by \cite{fksd01}.  \citet{fksd01} determine
the opening angle of the jets from the observed jet-break times,
assuming $\tau_j \propto \theta_j^{8/3}$ \citep[e.g.][]{sph99}, and
from this infer that there exists in GRBs a range in jet-opening
angles. \citet{fksd01} have taken a simple jet where the Lorentz
factor does not vary with angle (i.e. a non-structered jet) and the
observer is effectively looking straight at the center of the jet.
Correcting for the geometry of the explosion, gamma-ray energies then
appear narrowly clustered; the most energetic GRBs thus have the
narrowest jets. If we adopt this model, and combine this with our
result (Eqn.~\ref{t_jeqn}), $\tau_j \propto \Delta t \propto
\theta_j^{8/3}$, and using $\Delta t \propto 1/{\cal D}$,
we find
\begin{equation}
\theta_j \propto {\cal D}^{-3/8} ~.
\end{equation}
(where we have taken $\tau_j \propto \Delta t$).  Because relativistic
beaming is strongest along the line of sight, material moving directly
toward the observer will dominate the emission.  For an observer
located along the axis of the jet, i.e. $\theta_v \approx 0$, this
translates into $\theta_j \propto \gamma^{-3/8}$ (${\cal D} \propto
\gamma$ for $\theta_v = 0$ in Eqn.~\ref{deltaeqn}).  We thus find, in
this framework, that the fastests GRBs, with the highest $\gamma$,
have the narrowest jets.  A prediction of this model is that not only
should there be a variation in opening angles $\theta_j$ among bursts,
but they should be related to Lorentz factor $\gamma$ in this
proportion.  A progenitor model must predict that objects of a given
jet angle $\theta_j$ be produced with a probability $P \sim
P_{obs}/\theta_j^2 \propto \theta_j^{-4.54}$, where \citet{fksd01}
have shown that $P_{obs} \propto \theta_j^{-2.54}$. We note here that
\citet{ry01} have pointed out that the observed
distribution of angles may be biased; wide jets are less effective
in burning away the circumburst dust, and thereby will more often be
optically undetectable (dark) than narrow jets. 

This ``Many Morphologies'' scenario assumes that the observed
variation among GRBs is an intrinsic feature of the burst population,
i.e.~there exists a spectrum of GRB Lorentz factors (and corresponding
luminosities) and of jet sizes (wide and narrow opening angles). It is
unclear why there is such a broad range of afterglow opening angles
(ranging from 3 to more than 25 degrees; Frail et
al. [2001]\nocite{fksd01}).  Why, if bursts are all of roughly equal
total energy, should more luminous bursts such as GRB\,990123 tend to
have much narrower opening angles than less luminous bursts such as
GRB\,970508?  Also, why is there a dearth of very narrow opening
angles (i.e.~less than 3 degrees)?

In the next two sections we will assume that bursts are produced by
jets that are very similar in nature, i.e.~we will assume a generic
(single) morphology. Assuming, as discussed in Section
\ref{modelsection}, that the relation of Fig.~\ref{tjvtlag} has a
kinematic origin (i.e.~Eqn.~\ref{tdeltaeqn}), then the observed
variation among GRBs originates from variation in ${\cal D}$
(Eqn.~\ref{deltaeqn}) which depends on two variables; $\theta$ and
$\gamma$.  In more realistic jets two effects will likely play a role.
The first is where the variation in GRB properties is due to variation
of viewing angle, $\theta$, and, the second is where the variation in
GRB properties is due to the velocity structure of the jet,
i.e.~$\gamma$. In reality both viewing angle and velocity profile may
be important, in which case the combined effect needs to be taken into
account. This is beyond the scope of this paper.

\subsection{Single Morphology: Variation in Viewing Angle } \label{simplejet}

It was recently suggested by \citet{in01} that the lag-luminosity
relationship (Eqn. \ref{jpnlageqn}) could be explained by variation in
observer angle, $\theta_v$, from the axis of the jet. Here we have a
simple jet, where the Lorentz factor $\gamma = $ const.~for $\theta <
\theta_j$ and $\gamma = 0$ for $\theta > \theta_j$. However, now the
observer is not looking at the center of the jet, but is slightly
off-axis ($\theta_v \neq 0$). The lags derive from the difference in
time-of-flight of the near and far edges of an emitting internal shock
disk of finite size, $\Delta t \propto {\cal D}^{-1}$, and the
luminosity is defined as $L \approx $~const.~for ($\theta_v <
\theta_j$) and asymptotically varies as $L \propto {\cal D}^{3}$ for
($\theta_v > \theta_j$). \citet{in01} calculate that bursts observed
at angles $\theta_v$ a few times $\theta_j$ would peak in X rays
(possibly X ray rich GRBs) while for $\theta_v \sim \theta_j$ they
would peak in gamma rays. Using this model and $\theta_v \approx
\theta_j$, \citet{in01} were able to adequately reproduce the observed
lag-luminosity relationship (Eqn.~\ref{jpnlageqn}); this includes
GRB\,980425 \citep{gvv+98} which has very low luminosity, very large
spectral lag and very low variability.

\begin{figure}
\plotone{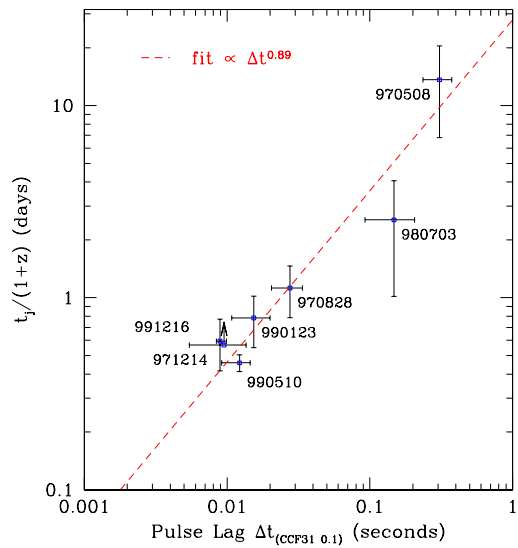} 
\caption{Plot of redshift-corrected burst pulse lags, $\Delta t$,
observed between BATSE channels 1 and 3, versus observed jet-break
times, corrected for redshift, $\tau_j \equiv t_j/(1+z)$.  Jet break
times $t_j$ are from \citet{fksd01} and pulse lags are from
\citet{nmb00}.  The fit, given by Eqn.~(\ref{t_jeqn}), does not include
GRB\,971214 which only has a lower limit on the jet-break time. 
\label{tjvtlag}}
\end{figure}

\begin{figure}
\plotone{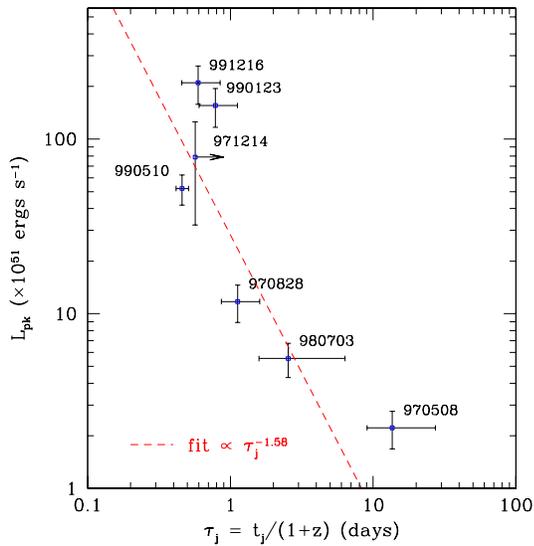} 
\caption{Plot of redshift-corrected burst peak luminosites $L_{pk}$,
versus redshift-corrected observed jet-break times $\tau_j \equiv
t_j/(1+z)$.  Jet break times $t_j$ are from \citet{fksd01} and
luminosities are calculated from \citet{jbp01}.  Because GRB\,971214
only has a lower limit on the jet-break time, it is not included in the
fit (given by Eqn.~\ref{Leqn}).
\label{Lvtlag}}
\end{figure}

\begin{figure}
\plotone{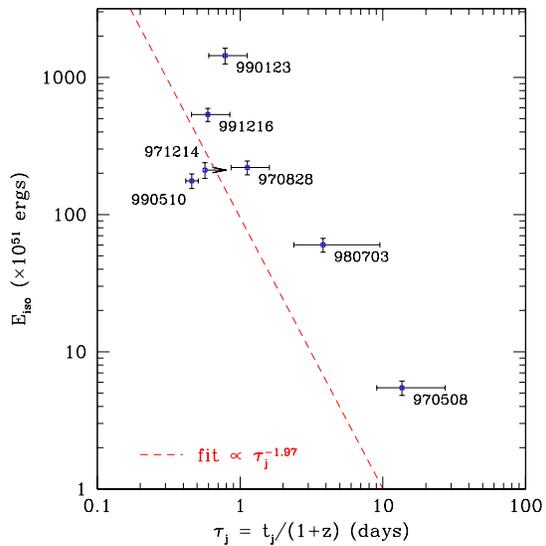} 
\caption{Plot of isotropic energy, $E_{iso}$, versus
redshift-corrected observed jet-break times $\tau_j$.  Jet break times
are from \citet{fksd01} and energies are from \citet{bfs01}.
GRB\,971214 had only a lower limit on the jet-break time and so is not
included in the fit (given by Eqn.~\ref{Eeqn}).
\label{Eisovtj}}
\end{figure}

The simplicity of this model allows for a demonstration of its
predicted adherence to the relationship of Eqn.~(\ref{t_jeqn}).  The
observed jet-break time, defined here as the time when $\gamma \equiv
1/\theta_j$, will occur at a radius $r_j \propto \gamma^{-2/3} \propto
\theta_j^{2/3}$ and thus will vary with observer angle $\theta_v$ from
the jet axis as
\begin{equation}
\begin{split}
\tau_j &\propto \frac{r_{j}}{c} (1 - \beta \cos [\theta_v - \theta_j]) \\ &\propto
	\theta_j^{2/3} (1- \beta \cos [\theta_v - \theta_j]) \\ &\propto \theta_j^{8/3}
	(1 + [(\theta_v - \theta_j)/\theta_j]^2), \quad \theta_v \geq \theta_j.
\label{tauj}
\end{split}
\end{equation}
Similarly the pulse lag will vary as
\begin{equation}
\Delta t = \frac{\Delta \tau }{{\cal D}} \propto \frac{(1 + [(\theta_v - \theta_j) \gamma]^2)}{\gamma}, \quad \theta_v \geq \theta_j~.
\label{deltat}
\end{equation}
Thus one sees that if $\gamma \approx 1/\theta_j$ one gets the relation
\begin{equation}
\Delta t \propto \tau_j
\label{simpletteqn}
\end{equation}
similar to Eqn. (\ref{t_jeqn}). In this model, the relation observed
in Fig.~\ref{tjvtlag} is a natural result of the range of off-axis
views, $\theta_v$, of a narrow jet.  Note that Eq. (\ref{tauj}) relates
to the afterglow phase, whereas Eq. (\ref{deltat}) relates to the GRB
phase. Two predictions can be made: {\it i)} $\gamma \approx
1/\theta_j$ to within a factor of a few lest the curve deviate from
the straight line of Fig.~\ref{tjvtlag}. So the afterglow lightcurve
breaks while moving at a substantial fraction of the original Lorentz
factor. {\it ii)} Since the ratio of maximum to minimum lag and that
of jet-break times shown in Fig.~\ref{tjvtlag} is about 30, and
assuming that $\theta_v$ is negligable for the bursts with the
shortest timescales (e.g.~GRB\,990510 and GRB\,971214), the implied
maximum timescale (in particular GRB\,970508) derives from a viewing
angle $\theta_{v(GRB\,970508)} \sim \sqrt{30}/\gamma \sim 6/\gamma$.
For example, letting $1/\theta_j = 2 \gamma/3$ and defining $\tau_j =
0.6 (1 + [2/3 \gamma (\theta_v - \theta_j)]^2)$ and $\Delta t = 0.01
(1 + [\gamma (\theta_v - \theta_j)]^2)$ over the range $0 < \gamma
(\theta_v -\theta_j) < 6$, one gets an acceptable fit to the data in
Fig. \ref{tjvtlag}.

A difficulty with this model is that $\theta_{v(GRB\,970508)}$ $ \sim
6/\gamma$ is viewed well outside the beaming angle $1/\gamma$ and thus
we expect the afterglow decay to be steeper than the $F_{\nu} \propto
t^{-1.2}$ that is typically observed.  That is, the afterglow decay
curve must already have broken.  We then also expect to observe a
steeper luminosity versus spectral lag curve
\begin{equation}
L_{pk} \propto {\cal D}^{\alpha+2} \approx {\cal D}^3 \propto (\Delta
	t)^{-3}
\end{equation}
\citep[e.g.][]{jay01} than is observed (see Eqns.~\ref{jdslageqn} \&
~\ref{jpnlageqn}).  \citet{in01} mention future work that will show
that the break in the GRB spectrum will change the angular dependence
such that $\theta_{v(GRB\,970508)} \sim \sqrt[4]{30}/\gamma \sim
2.5/\gamma$ which is a more reasonable range.  Such detailed
calculations are beyond the scope of this paper.

Bursts with a given perceived Doppler factor, ${\cal D}$, will have
probabilities of detection that scale with luminosity $L_{pk} \propto
{\cal D}$ (Eqns.~\ref{deltaeqn} \& \ref{lumvsdeltaeqn}), and linearly
with angle, $\theta_v$, 
\begin{equation}
P_{obs} \propto L_{pk} \theta_v \propto {\cal D} \theta_v \propto
\frac{\theta_v}{1 + [(\theta_v-\theta_j) \gamma]^2}, \quad \theta_v
\geq \theta_j~.
\label{probeqn_singleburst}
\end{equation} 
The distribution has a maximum at the jet-opening angle $\theta_v \sim
\theta_j$.  For $\theta_v < \theta_j$ the observed jet opening angle
will just be $\theta_j$, so there will be no observations of smaller
opening angles.  Observations show no bursts below an angle 0.05, so
we would identify $\theta_j \lesssim 0.05$ \citep{fksd01}.  The
probability will asymptotically decrease as $P_{obs} \propto
\theta_v^{-1}$.  In \citet[see their Fig. 1]{fksd01}, for the 15
bursts with known redshift and jet-break time, it is found that
$P_{obs} \propto \theta_j^{-2.54}$ which is consistent with
Eqn.~(\ref{probeqn_singleburst}) to the precision of the sparse data.
Thus we find qualitative agreement with the observed probability
distribution of bursts.

\subsection{Single Morphology: A Structured Jet } \label{structuredjet}

In the following we will consider a jet with a velocity profile,
$\gamma = \gamma(\theta_v)$. Because relativistic beaming is strongest
along the line of sight, material moving directly toward the observer
will dominate the emission. So in the following we will take it as
sufficient to consider a small region around $\theta_v$ ($|\theta -
\theta_v| < 1/\gamma$). Within this narrow cone ${\cal D} \propto
\gamma$, and $\Delta t \propto 1/\gamma$ (Eqn.~\ref{deltaeqn}).  This
naturally satisfies the lag-luminosity relation
(Eqns.~\ref{jpnlageqn},\ref{jdslageqn}).

The following discussion will be qualitative. The model begins as a
wide jet that emerges from the GRB source so that the fastest moving
material moves along the narrow core of the jet and progressively
slower material moves along increasing inclination angles from the jet
core.  This situation might be seen as a narrow, fast jet, surrounded
by a slower, wider ``coccoon'' of material.  The GRB arises from this
hot, fast moving material. After the GRB phase there is a shock that
moves into the interstellar medium (ISM) that roughly maintains the
velocity profile of the original jet.  The extent of the jet is
sufficiently large ($>$ 30 degrees) that the actual geometry of the
jet, i.e., seeing the `edge' of the jet \citep{sph99} does not occur
(except for very off-edge observed GRBs; possibly GRB\,980425
[Salmonson 2001]\nocite{jay01}). The jet break in afterglow
lightcurves is then predominantly from sideways expansion of the jet
\citep{sph99,rhoads99}.

The shock at low inclination angle from the jet axis leads the shock
at progressively higher inclination angle.  This is a very plausible
situation in the collapsar model (but does not require a collapsar
model: we only require a velocity profile with slower material at
higher inclination angles). In the collapsar model the core of the jet
will break through the stellar surface first, followed by a widening
and sideways expansion as the jet becomes fully established.
Numerical simulations by \citet{amimm00} indicate that this may happen
on the order of seconds or more.  Thus the core of the afterglow will
be ahead of the wings by a few light-seconds and this lead will be
maintained throughout most of the evolution due to the high velocity
($v \sim c$) of the ejecta.  This ``arrowhead'' jet structure will
cause unbalanced sideways expansion of causally connected shocked
material.  As ever larger and wider concentric rings of shocked
material become causally connected, larger and larger regions will
begin to expand sideways.  This progression of expansions yields ever
later jet-break times for observers at higher inclination angles.
Details of this evolution will require numerical modeling of as yet
poorly understood hydrodynamics of collisionless shocks. \citet{dg01}
modeled an afterglow jet with a $\gamma \propto \theta_j^{-q}$
structure, but did not allow sideways spreading of the jet.

\section{Discussion}

Herein we have presented a tight relation between GRB pulse lags and
afterglow jet-break times (Eqn.~\ref{t_jeqn}). As spectral lag was
shown to be related to the isotropic gamma-ray peak luminosity
\citep{nmb00} this also represents a correlation between peak
luminosity and jet-break time.  This is, to our best knowledge, the
first known direct correlation between a property of the gamma-ray
burst phase (the pulse lag or peak luminosity) and the afterglow phase
(the jet-break time).  As observed timescales (variability or spectral
lags) as well as peak luminosity naturally have a strong dependence on
the Lorentz factor, $\gamma$, or angle with respect to the motion of
the outflow, $\theta$, we propose that the variety among GRBs has a
purely kinematic origin.  The emergence of simple trends between
jet-break time, pulse lag and luminosity gives us clues and
constraints on how such a model can be constructed.

\citet{fksd01} infer that the differences among observed GRB energies
are due to a range of jet opening angles. Within this framework we
have then shown that the fastest GRBs, with the highest $\gamma$, have
the narrowest jets (Sec.~\ref{burstpop}). A prediction of this model
is that not only should there be a variation in opening angles
$\theta_j$ among bursts, but they should be related to Lorentz factor
$\gamma$ as $\theta_j \propto \gamma^{-3/8}$, and jet angles
$\theta_j$ have to be produced with a probability $P \sim
P_{obs}/\theta_j^2 \propto \theta_j^{-4.54}$.

We then have explored the possibility that variation among GRBs, all
now assumed to be morphologically the same, is caused by variation in
observer viewing angle from the jet axis (Sec.~\ref{simplejet}), or
Lorentz factor of the jet as a function of angle from the jet axis
(Sec.~\ref{structuredjet}).  These models qualitatively show that
variation of the observed Doppler factor, ${\cal D}$,
(Eqn. \ref{deltaeqn}) from a single jet morphology can produce the
observed variations among GRB luminosities and timescales: $L_{pk}
\propto N_{pk} \propto 1/\Delta t \propto 1/\tau_j \propto {\cal D}$.
These models also provide a natural explanation for the probability
distribution of observation angles $P_{obs}$ consistent with that
reported by \citet{fksd01}.  A realistic model of a single-burst
morphology will have both perspective effects as in Section
\ref{simplejet} and jet-structure effects as in Section
\ref{structuredjet}.

Set inside the collapsar progenitor model \cite[e.g.][]{mcfw99} a
cohesive picture begins to emerge.  A jet driven out from the center
of the star will vary as a function of angle from the jet axis.
Ejecta at the center will be faster and lighter and, bearing the brunt
of the as yet unknown acceleration mechanism, will be more
fractured. At larger angles from the jet core, ejecta will interact
with the stellar wall, and thus will be slower and will entrain more
baryons.  Thus one expects to see high luminosities, short pulse lags
and high variability as well as an early jet break time for bursts
viewed on axis, while higher viewing inclinations will yield lower
luminosities, longer pulse lags, smoother bursts and later jet break
times.  Thus the variability-luminosity relationship \citep{rlfr+01}
as well as the spectral lag-jet break time and the spectral
lag-luminosity (Eqns.~\ref{t_jeqn} \& \ref{jpnlageqn}) can be
naturally accomodated.

\acknowledgements This work was performed under the auspices of the
U.S. Department of Energy by University of California Lawrence
Livermore National Laboratory under contract W-7405-ENG-48. TJG
acknowledges support from the Sherman Fairchild Foundation. We wish to
thank Daniel Reichart and Dale Frail for useful comments.






\end{document}